# AlN Gate Interlayer for UWBG AlGaN Transistors with Breakdown Field >6.9 MV/cm and PFOM >1.8 GW/cm$^2$

Seungheon Shin[1,a)], Jonathan Pratt[1], Joe McGlone[1], Yinxuan Zhu[1], Brianna A. Klein[3], Andrew Armstrong[3], Andrew A. Allerman[3], and Siddharth Rajan[1,2]

[1]*Department of Electrical & Computer Engineering, The Ohio State University, Columbus OH 43210, USA*
[2]*Department of Materials Science & Engineering, The Ohio State University, Columbus OH 43210, USA*
[3]*Sandia National Laboratories, Albuquerque, New Mexico 87123, USA*

**Abstract— We report the demonstration of regrown epitaxial AlN gate interlayers with ultra-wide bandgap (UWBG) AlGaN polarization-graded field effect transistors (PolFETs). The introduction of the epitaxial AlN gate interlayer enables significant improvement in breakdown strength, with average breakdown field exceeding 6.94 MV/cm, which represents state-of-the-art for lateral field effect transistors, while maintaining excellent on-state current density exceeding 1 A/mm. The integration of epitaxial AlN enables state-of-the-art power-switching figure of merit exceeding 1.87 GW/cm$^2$ at a breakdown voltage exceeding 1.45 kV. This work shows the potential of UWBG AlGaN for next-generation high-power switching and RF applications with enhanced device performance established by a high-quality epitaxially regrown AlN gate interlayer.**

[a)] Authors to whom correspondence should be addressed

Electronic mail: *shin.928@osu.edu*

Ultra-wide bandgap (UWBG) AlGaN, also referred to as high Al-content or Al-rich AlGaN, has demonstrated outstanding device performance in recent years, including multi-kV breakdown voltage [1-6], large average breakdown field [5-8], high current drive [6, 9, 10, 27], high-temperature operation [11, 12, 31], and promising RF performance [10, 13, 14], positioning UWBG AlGaN as a strong candidate for next-generation high-power switching and mm-wave applications. These advances are important for improving the Johnson figure of merit (JFOM) [32] and power-switching figure of merit (PFOM), which guide device design for RF and power applications, respectively. While materials figures of merit provide an estimate of the ultimate performance limits, realizing these limits in devices is a challenge as breakdown fields of materials approach extreme values such as in ultra-wide bandgap transistors.

In RF transistors, the theoretical maximum output power density [33] under Class A operation can be estimated as

$$\frac{P_{OUT}}{W} = \frac{(V_{MAX}-V_{KNEE})\, I_{MAX}}{8} \approx \frac{v_{sat} F_{BR} I_{MAX}}{2\pi f_T} \quad (1)$$

where $V_{MAX} - V_{KNEE} \approx V_{BR} = F_{BR} L_{GD}$, $f_T \sim \frac{v_{sat}}{L_{GD}}$, and $V_{BR}$, $F_{BR}$, $L_{GD}$, and $v_{sat}$ are breakdown voltage, average breakdown field, gate-to-drain spacing, and saturation velocity, respectively. In devices targeted towards RF power amplification, it is important to achieve a high $F_{BR} I_{MAX}$ product. In power devices, total power loss can be approximated [34] as

$$P_{Loss} = P_{SW} + P_{COND} \propto \frac{1}{\sqrt{F_{BR}{}^2 \mu_n}} \quad (2)$$

where $P_{SW}$, $P_{COND}$, and $\mu_n$ are switching loss, conduction loss, and electron mobility. This relation shows that realizing a high breakdown field within a device is critical for minimizing the conduction and switching losses. In summary, the realization of high breakdown field in a device while simultaneously maintaining high current density and mobility are pre-requisites for achieving high-performance ultra-wide bandgap RF and power transistors.

In a previous report, Zhu et al. demonstrated a cutoff frequency above 85 GHz in UWBG AlGaN transistors [10], enabled by low-contact-resistance engineering [15]. Furthermore, previous work established the high-voltage capability of UWBG AlGaN by demonstrating breakdown robustness exceeding 2 kV together with high PFOM in heterostructure field-effect transistors (HFETs) and polarization-graded field-effect transistors (PolFETs) [1, 2, 4-6]. However, further progress toward the theoretical material limits of UWBG AlGaN will require concurrent advances in high-quality gate insulators, advanced field-management schemes for extreme electric fields, significant reduction in access and contact resistance, improved passivation to minimize switching dispersion, and fundamental understanding of carrier mobility in UWBG AlGaN. Among these, gate insulator quality is particularly important because it directly affects gate leakage, interfacial charge control [7, 16], and electrostatic modulation of the channel, while also governing susceptibility to premature breakdown under the extreme electric fields in UWBG AlGaN. In conventional Schottky-gated structures, large gate current can arise before the intrinsic material limit is reached, resulting in premature breakdown well below the theoretical breakdown electric field (>10 MV/cm) in lateral UWBG AlGaN transistors.

To address these interrelated challenges, we investigate here UWBG AlGaN PolFETs integrating epitaxially regrown AlN gate interlayers. The AlN-integrated devices show an extreme average breakdown field exceeding 6.94 MV/cm while maintaining high $I_{MAX}$ above 1 A/mm, and a state-of-the-art PFOM above 1.87 GW/cm² in lateral power transistors showing a similar kV-class (~ 2 kV) breakdown voltage range. These results establish a gate insulator engineering route for UWBG AlGaN targeting next-generation mm-wave and high-power switching applications.

The starting epitaxial structures employed in this work were grown on AlN substrates in a TNSC-4000HT MOCVD reactor. The stack comprises a 2 nm heavily doped back-barrier, a 30 nm unintentionally doped (UID) $Al_{0.5}Ga_{0.5}N$ buffer, a 10 nm AlGaN channel graded from $Al_{0.5}Ga_{0.5}N$ to $Al_{0.8}Ga_{0.2}N$ toward spacer, a 30 nm n-$Al_{0.8}Ga_{0.2}N$ spacer, and 32.5 nm reversed-graded n++ AlGaN contact layers. The layer sequence, thicknesses, alloy compositions, and doping conditions are shown in Fig. 1(a). Detailed functions of each layer and the associated epitaxial design considerations are described in our previous work [6]. The device fabrication began with low-damage $Cl_2$-based ICP-RIE to remove the contact layers in the access regions, followed by 7.5 nm plasma-assisted molecular beam epitaxy (PAMBE) AlN epitaxial regrowth (samples referred to here as "AlN Interlayer"). The regrown AlN layers in the ohmic regions were etched using the same low-damage ICP-RIE process, followed by non-alloyed ohmic metal evaporation (Ti/Al/Ni/Au = 20/120/30/100 nm), and mesa isolation (~ 240 nm) using ICP-RIE. Finally, Ni/Au (30/200 nm) gate metal was deposited. All processing flow was done using direct-write optical lithography. AlN regrowth and etching for ohmic regions steps were omitted for the control sample (samples referred to here as "No AlN Interlayer"). The simulated ideal energy band diagrams and electron distributions under the gate region for both devices are presented in Fig. 1(b), (c). The simulated electron profile shows extra charge accumulation in the AlGaN layers due to reduced surface depletion induced by the additional polarization charges in the epitaxially regrown AlN layer. The energy band diagram for ohmic regions and high-resolution X-ray diffraction for the epitaxial layer template are described in [6].

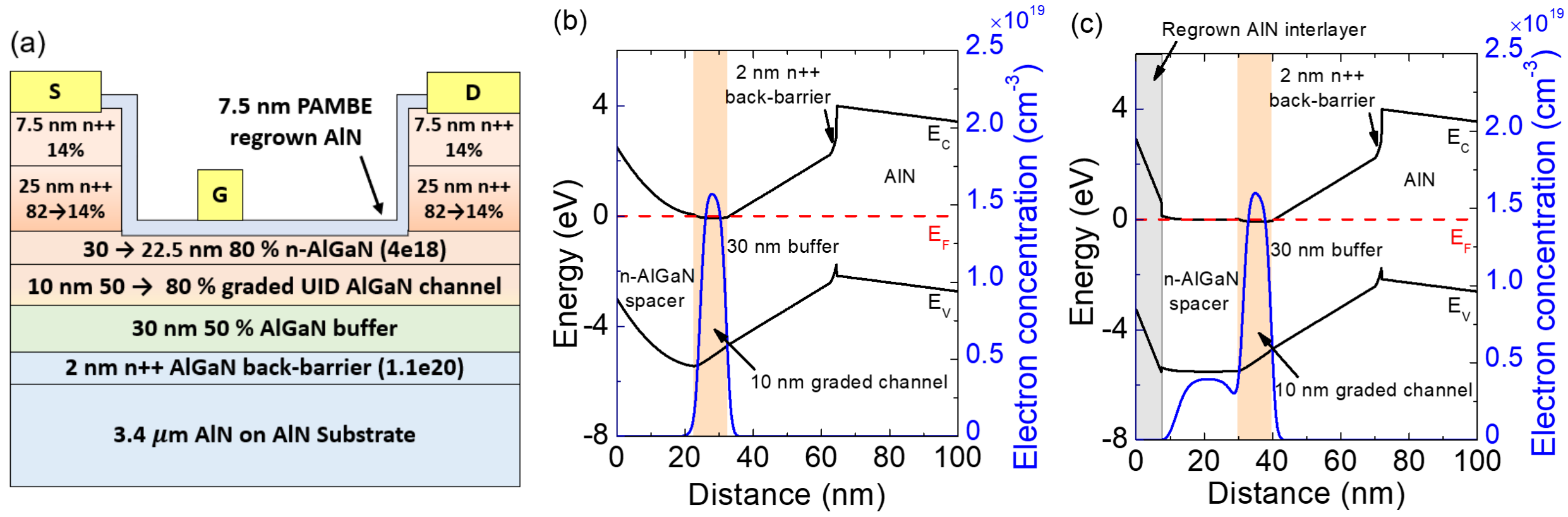

Figure 1. (a) Schematic of epitaxial and device structures with 7.5 nm AlN interlayer, 1-D Schrodinger simulated ideal energy band diagrams and electron distribution profiles under the gate region for (b) No AlN interlayer devices, (c) AlN interlayer devices

The Hall measurements and transmission line measurement (TLM) results for each device structure are summarized in Table 1. The Hall measurement results indicate that sheet resistance ($R_{SH}$) decreased by about 6 % with the AlN interlayer, compared with the No AlN interlayer sample, mainly due to increased charge density, whereas the mobility variation was negligible. It is suggested that the additional polarization charge in the regrown AlN layer compensates for the surface depletion resulting in increased charge density. TLM results showed only minor variation in ohmic contact properties, indicating the process feasibility of regrown AlN integration with AlN etch-back process.

Furthermore, the charge boosting effect was also confirmed by capacitance-voltage (C−V) measurements. The double sweep C−V measurements were done up to the forward turn-on voltage for each device (Fig. 2(a)). From the extracted electron profiles, a second electron peak was observed near the surface in the AlN interlayer device, while positive bias could not be applied to the No AlN interlayer devices due to high conductivity (Fig. 2(b)). This result suggests that regrown AlN integration is effective in generating additional conductive charge, leading to improved on-state device performance.

Table 1. Hall measurement and TLM results

| | No AlN interlayer | AlN interlayer |
|---|---|---|
| Sheet resistance (kΩ/□) | 2.36 | 2.25 |
| Electron mobility ($cm^2$/V·s) | 222 | 219 |
| Charge density (× $10^{13}$ $cm^{-2}$) | 1.19 | 1.26 |
| Contact resistance (Ω·mm) | 2.24 | 2.07 |
| Contact resistivity (× $10^{-5}$ Ω·$cm^2$) | 2.05 | 1.96 |

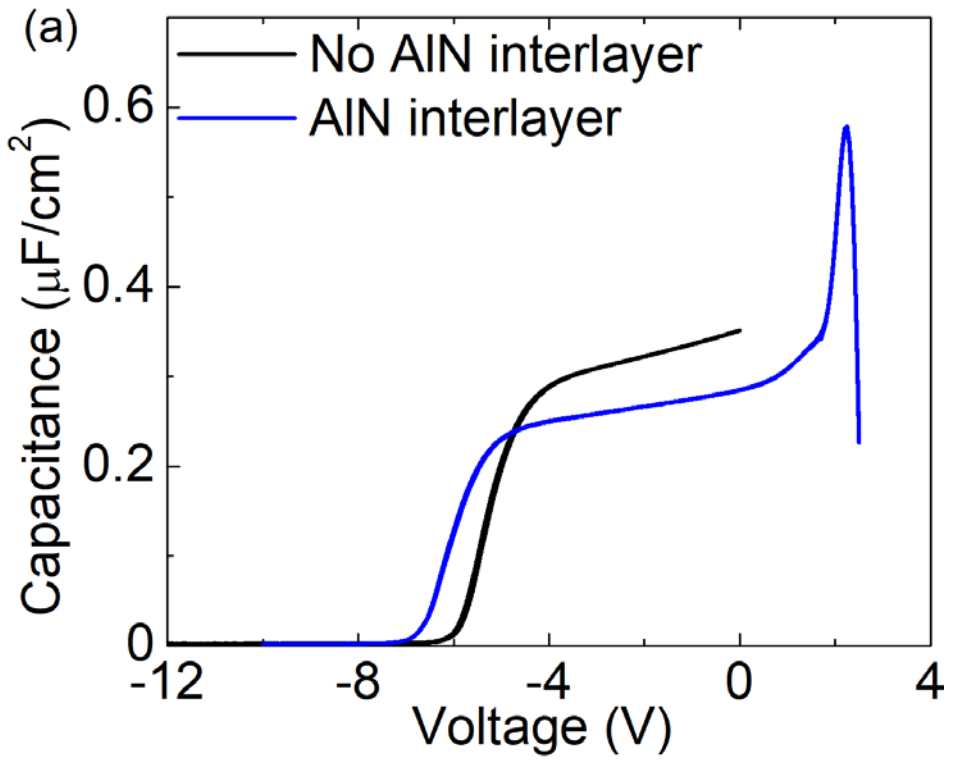

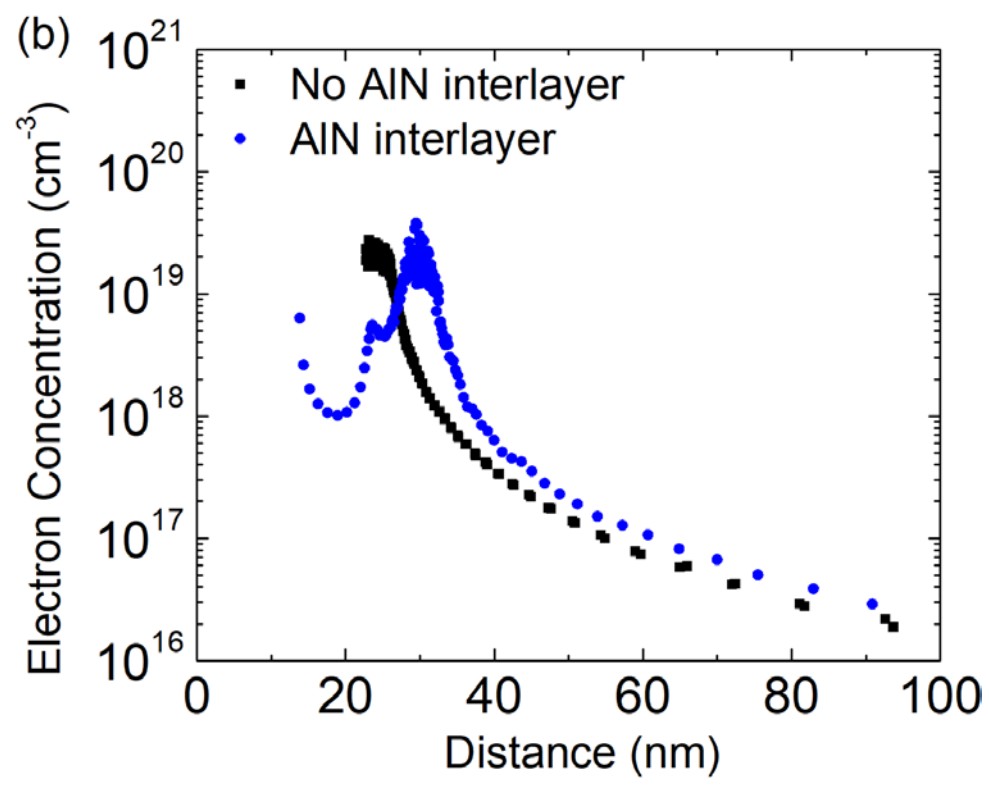

Figure 2. (a) Double sweep C−V measurements up to forward turn-on voltage (b) extracted electron profile from C−V characteristics

For device characteristics, three typical devices varying gate-to-drain spacings are reported here. For No AlN interlayer devices, $L_{GD}$ = 0.82, 4.06, and 6.84 μm with nominally same $L_G$ = 0.9 μm and $L_{SG}$ = 0.6 μm. For AlN interlayer devices, $L_{GD}$ = 0.86, 4.03, and 6.96 μm with nominally same $L_G$ = 1.24 μm and $L_{SG}$ = 0.34 μm. Among the shortest gate-drain spacing devices, the AlN interlayer device ($L_{GD}$ = 0.86 μm) exhibited a maximum on-state current ($I_{MAX}$) of 1.02 A/mm at $V_{GS}$ = 3 V, $V_{DS}$ = 20 V (Fig. 3(a)), while the No AlN interlayer devices showed 0.9 A/mm under the same bias condition (Fig. 3(b)), which agrees with the trend expected from the Hall/C−V characterization. In addition, AlN interlayer devices exhibited a 7 × lower gate leakage current than No AlN interlayer devices due to the higher barrier height at gate-metal/AlN interface. This corresponds to on/off-ratios of 5.4× $10^8$ and 6.4 × $10^7$, for AlN interlayer and No AlN interlayer devices, respectively (Fig. 3(c)). These results suggest that thin AlN epitaxial regrowth can simultaneously improve both on-state and off-state as a insulating gate interlayer through increased charge density and higher gate tunneling barrier height.

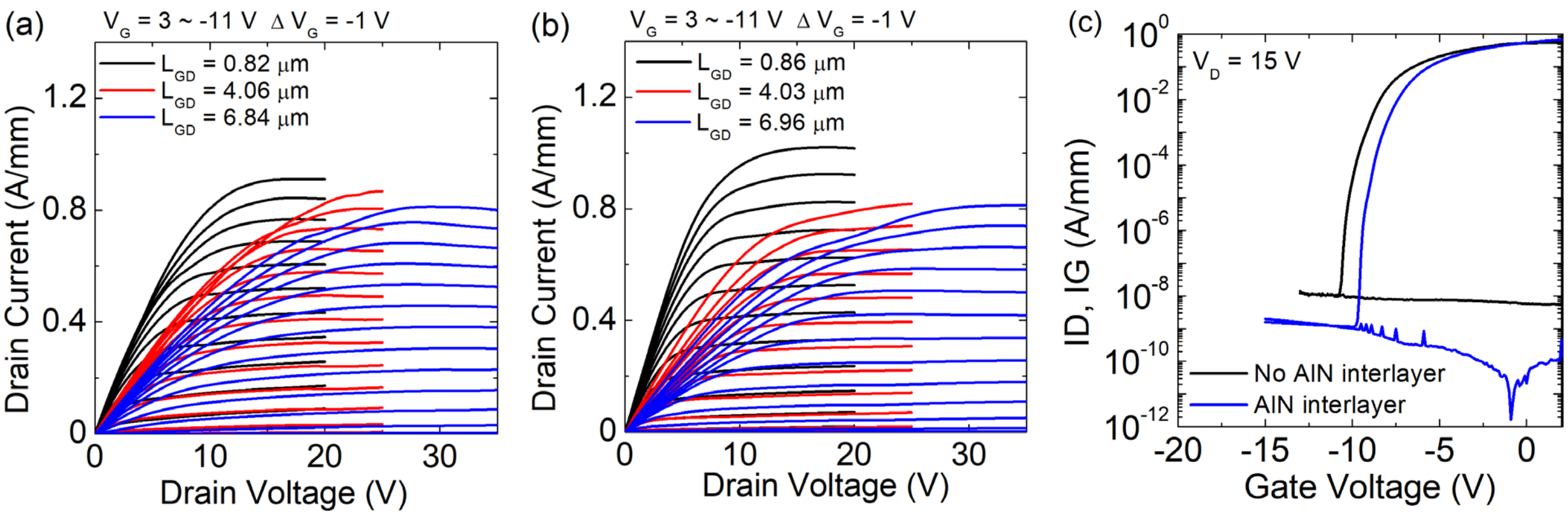

Figure 3. DC output characteristics with different $L_{GD}$ for (a) No AlN interlayer (b) AlN interlayer devices, (c) transfer curve for both devices, solid lines: drain current, dashed lines: gate current

Small-signal measurements were done for AlN interlayer device with $L_{GD}$ = 0.86 μm, using Maury Microwave MT2000 system. Fig. 4 shows the representative current gain, maximum unilateral gain (MUG) for the valid region only, and maximum stable gain (MSG) measured at $V_{GS}$ = -4.6 V, $V_{DS}$ = 20 V. The extracted cutoff frequency ($f_T$) and maximum oscillation frequency ($f_{MAX}$) were 5.6 GHz and 16 GHz, respectively. It should be noted that the RF performance is mainly limited by relatively high contact resistance.

The trap-related characteristics were investigated through pulsed I−V measurements for both No AlN interlayer and AlN interlayer devices with the smallest $L_{GD}$. The measurements were done using Keithley 4200A. The pulsing condition for both devices was set to be pulse width = 6 μs, duty cycle = 0.1 %, with $V_{DSQ}$ = 30 V, and the $V_{GSQ}$ was more negative bias than pinch-off voltage ($V_P$) to include the effect of deeper traps below channel layer ($V_{GSQ} \sim V_P - 5$ V). The current drops in pulsed I−V measurements were around 28 % and 19 % from DC $I_{MAX}$ for No AlN interlayer and AlN interlayer devices, respectively (Fig. 5(a), (b)).

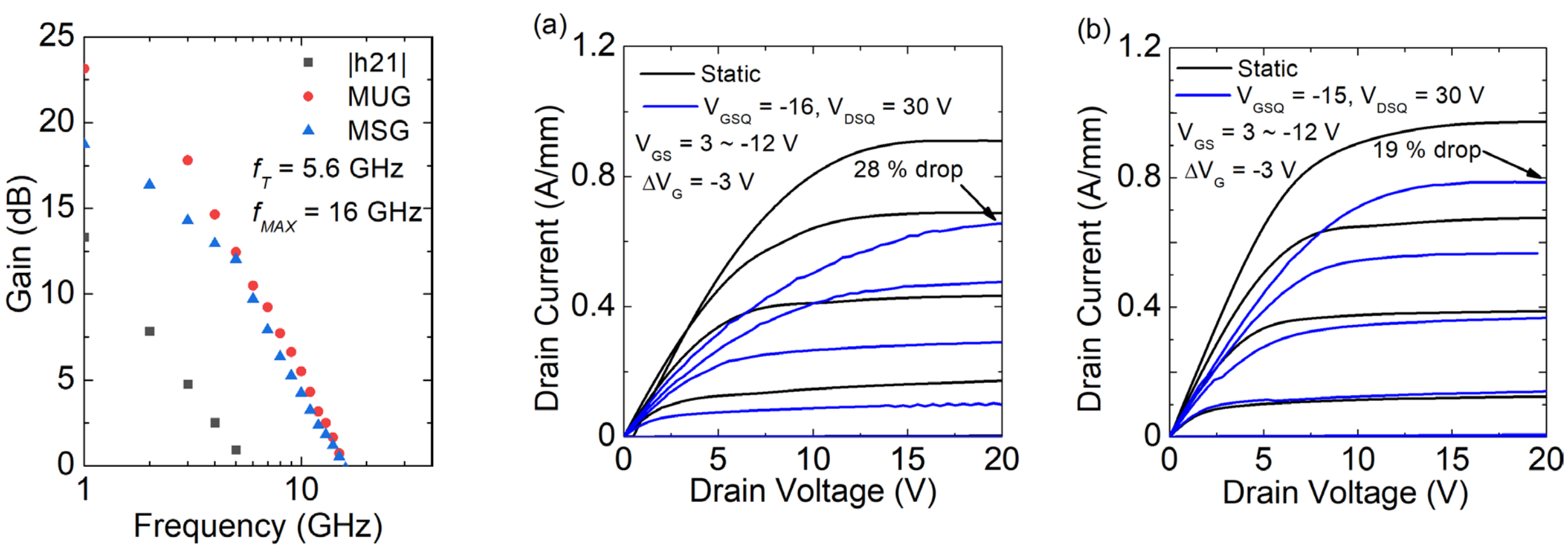

Figure 4. RF characteristics for AlN interlayer devices with $L_{GD}$ = 0.86 μm measured at $V_{GS}$ = -4.6 V, $V_{DS}$ = 20 V

Figure 5. Pulsed I−V measurements at $V_{GSQ} \sim V_P - 5$ V, $V_{DSQ} = 30\ V$ for (a) No AlN interlayer devices (b) AlN interlayer devices

The off-state characteristics were evaluated using three-terminal high-voltage breakdown measurements with Keysight B1505A analyzer. The breakdown voltage ($V_{BR}$) here is defined at the drain voltage where the off-state ($V_{GS}$

~ $V_P$ – 5 V) drain current exceeds 1 mA/mm. Typical breakdown measurements and the dependence of breakdown voltage and breakdown field on gate-drain spacing are shown in Figure 6 (a) and (b), respectively.

The AlN interlayer devices with $L_{GD}$ = 0.86 μm exhibited $V_{BR}$ = 600 V, corresponding to 6.94 MV/cm of average breakdown field (using the definition $F_{BR} = V_{BR}/L_{GD}$). The $V_{BR}$ for longer $L_{GD}$ (4.03 and 6.96 μm) was 1.45 and 1.67 kV, respectively, but with a lower average breakdown field. For No AlN interlayer devices showed significantly lower corresponding breakdown fields of $F_{BR}$ (~ 3.1 MV/cm at $L_{GD}$ = 0.82 μm). These results suggest that replacing the surface material at gate edge with a higher critical field material can improve the average breakdown field in devices. Two-dimensional device simulations were done using Silvaco TCAD [35]. Fig. 7(a) shows the two-dimensional contour of the x-direction electric field at a drain voltage of 600 V with $L_{GD}$ = 0.86 μm. The x-direction electric field distribution at gate-metal/regrown AlN interface in breakdown condition is shown in Fig. 7(b), suggesting a peak electric field magnitude of ~ 16 MV/cm, which is close to the expected breakdown field of AlN [36, 37]. However, it should be noted that the simulated peak electric field may be overestimated due to the absence of the correct definition for surface and interface traps in the TCAD simulation, and the sharp corners assumed in the simulations. In reality, the field peak is expected to be lower and more spread out.

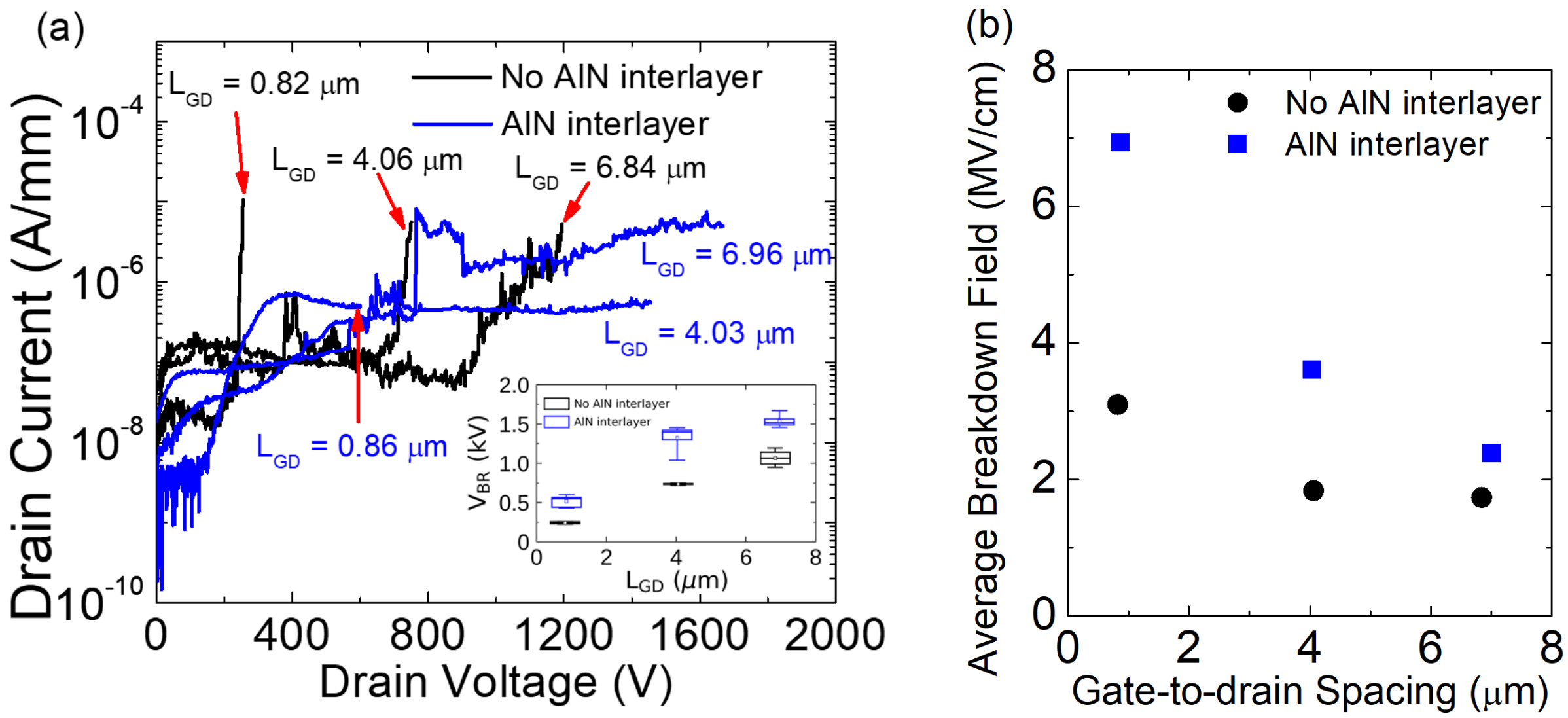

Figure 6. Three-terminal breakdown measurements with different $L_{GD}$, where blue and black lines represent AlN interlayer and No AlN interlayer, respectively. (b) $L_{GD}$-dependent $F_{BR}$,

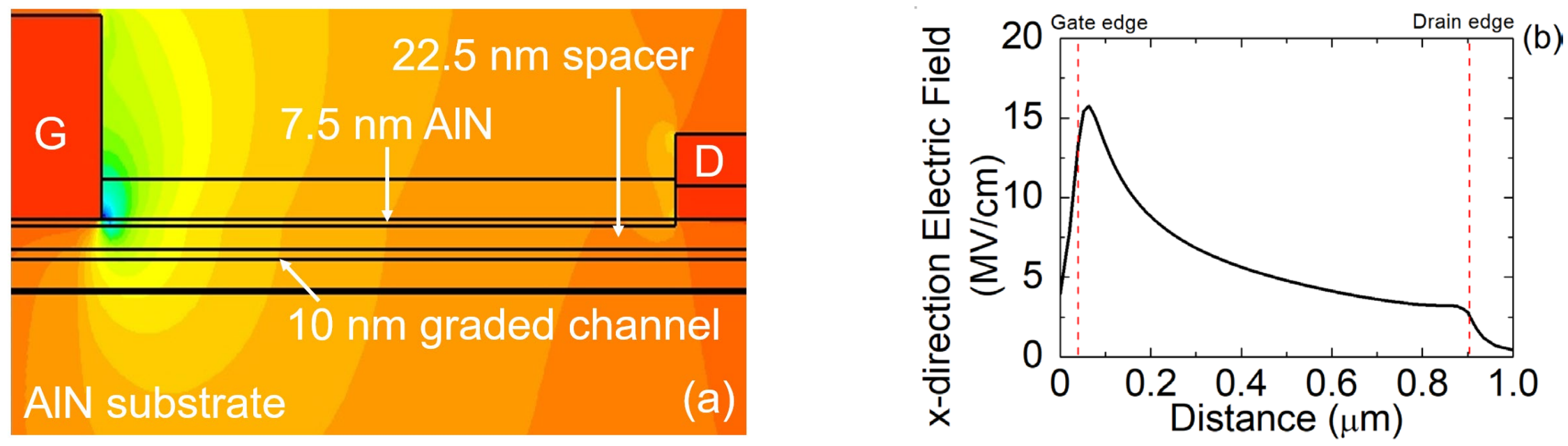

Figure 7. TCAD simulation with AlN interlayer device with $L_{GD}$ = 0.86 μm under breakdown condition (a) contour plot of x-direction electric field, (b) Electric field distribution at gate-metal/regrown AlN interface

From the linear region of output characteristics, the specific on-resistance ($R_{on.sp} = (2L_T + L_{SD}) \times R_{ON}$) of 2.12 and 2.09 m$\Omega\cdot$cm$^2$ were obtained in the longest $L_{GD}$ for No AlN interlayer and AlN interlayer, respectively (Fig. 3(a), (b)). Based on the $R_{on.sp}$ and breakdown results, the power-switching figure of merit (PFOM $= V_{BR}^2/R_{on.sp}$) was estimated to be ~ 1.87 GW/cm$^2$ for AlN interlayer devices, while No AlN interlayer devices showed ~ 0.67 GW/cm$^2$ (Fig. 8) [1, 5, 6, 17-25]. This represents the highest power switching figures of merit among WBG lateral transistors, including other semiconductors such as GaN, $Ga_2O_3$.

Furthermore, device performance comparison in state-of-the-art UWBG AlGaN transistors (Al composition above 40 % in channel) for maximum on-current and average breakdown field is presented in Fig. 9 [5-8, 14, 26-30]. These results suggest that epitaxially regrown AlN gate interlayer integration provides significant improvements in both on-state and off-state device performance as well, exhibiting record-high combinations of average breakdown field and maximum on-current.

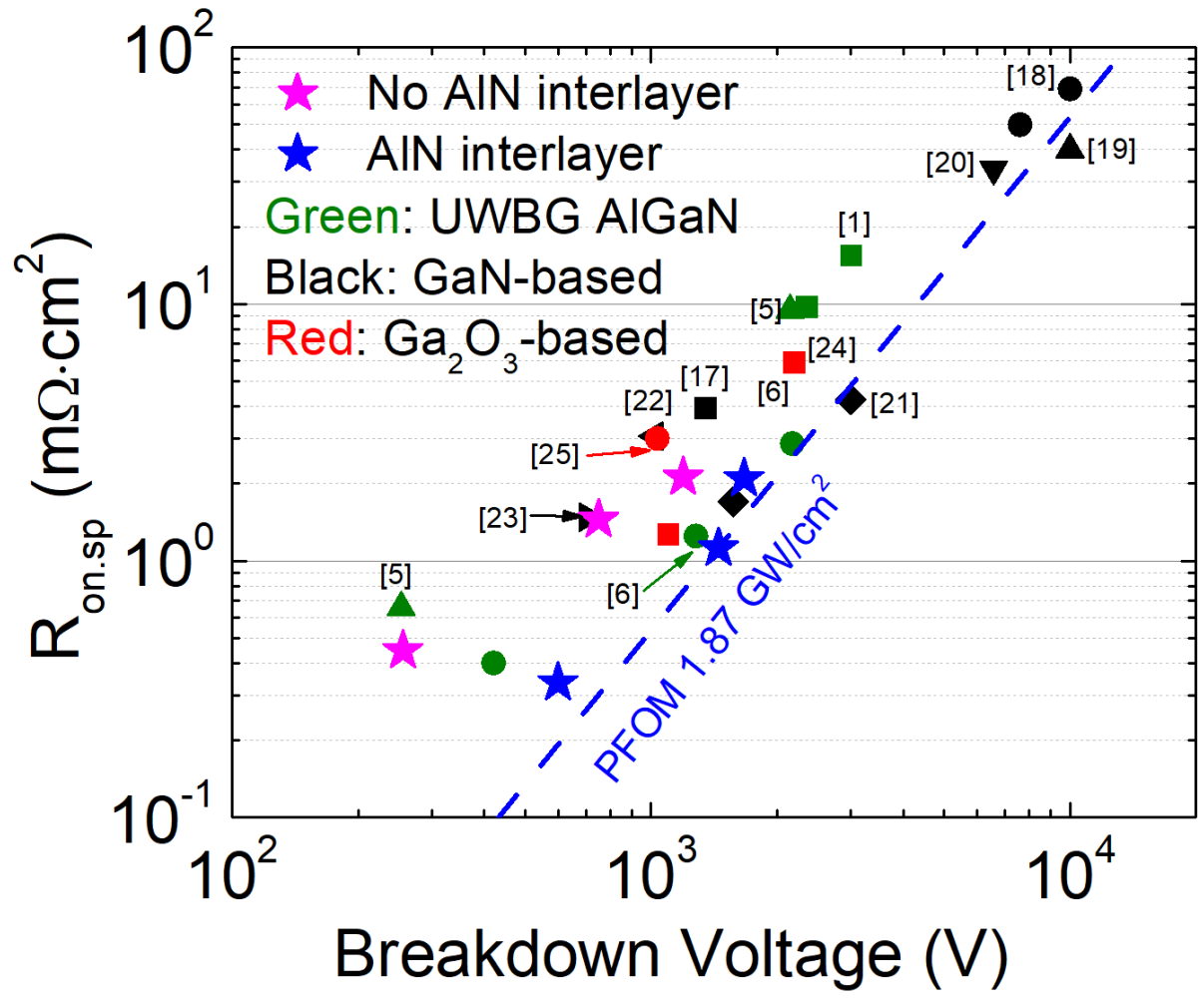

Figure 8. $R_{on.sp}$ vs. $V_{BR}$ benchmark in the state-of-the-art lateral power device materials [1, 5, 6, 17-25].

Figure 9. $F_{BR}$ *vs.* $I_{MAX}$ benchmark plot in UWBG AlGaN channel transistors (channel Al % >40 %) [5-8, 14, 26-30]

In conclusion, we report on the demonstration of PAMBE-regrown AlN gate interlayer integration on UWBG AlGaN PolFET platforms. A high average breakdown field (>6.94 MV/cm) in lateral field effect transistors was achieved with high on-state current density of 1.02 A/mm in devices with an AlN gate-insulating interlayer. A promising power-switching figure of merit (PFOM ~ 1.87 GW/cm$^2$) was realized in devices with $L_{GD}$ = 4.03 μm ($V_{BR}$ = 1.45 kV). This gate interlayer integration redefines UWBG AlGaN for next-generation mm-wave and high-power-switching applications.

This work was funded by ARO DEVCOM under Grant No. W911NF2220163 (UWBG RF Center, program manager Dr. Tom Oder). This article has been co-authored by employees of National Technology & Engineering Solutions of Sandia, LLC under Contract No. DE-NA0003525 with the U.S. Department of Energy (DOE). The employee owns all right, title and interest in and to the article and is solely responsible for its contents. The United States Government retains and the publisher, by accepting the article for publication, acknowledges that the United

States Government retains a non-exclusive, paid-up, irrevocable, world-wide license to publish or reproduce the published form of this article or allow others to do so, for United States Government purposes. The DOE will provide public access to these results of federally sponsored research in accordance with the DOE Public Access Plan https://www.energy.gov/downloads/doe-public-access-plan. SNL is managed and operated by NTESS under DOE NNSA contract DE-NA0003525

**Author Declarations**

**Conflict of Interest**

The authors have no conflicts to disclose.

**Data Availability**

The data that support the findings of this study are available within the article.

**Supplementary materials**

The supplementary material includes transmission line measurements comparison between the structures with and without contact layer etching, improvements on breakdown robustness enabled by gate-connected field-plate structures, and the $L_{GD}$-dependent breakdown field trends.